# Substantial magneto-electric coupling near room temperature in $Bi_2Fe_4O_9$

A. K. Singh, S. D. Kaushik, Brijesh Kumar, and S. Patnaik

School of Physical Sciences, Jawaharlal Nehru University, New Delhi 110067, India

P. K. Mishra

Technical Physics & Prototype Engineering Division, Bhabha Atomic Research Centre, Mumbai 400085, India

A. Venimadhav

Cryogenic Engineering Center, Indian Institute of Technology, Kharagpur 721302, India

V. Siruguri

UGC-DAE-Consortium for Scientific Research Mumbai Centre,

Bhabha Atomic Research Centre, Mumbai – 400085, India

# Abstract

We report remarkable multiferroic effects in polycrystalline $Bi_2Fe_4O_9$. High-resolution X-ray diffraction shows that this compound has orthorhombic structure. Magnetic measurements confirm an antiferromagnetic transition around 260 K. A pronounced inverse S-shape anomaly in the loss tangent of dielectric measurement is observed near the Néel temperature. This feature shifts with the application of magnetic field. These anomalies are indicative of substantial coupling between the electric and magnetic orders in this compound.

The recent surge in the search for multiferroic materials owes its origin to the discovery of magneto-electric effect, over a century ago, by Pierre Curie [1] and its material verification in $PbFe_{0.67}W_{0.33}O_3$ by Smolenskii [2]. By definition, these materials can sustain induced magnetization with the application of electric field and electric polarization by means of magnetic field and thus have a huge spread of potential applications such as non-volatile ferroelectric memories, sensors and actuators. This magneto-electric interdependence occurs because of pronounced interplay between various subsystems of crystal structure that lead to coupling between the ordered ferroelectric and (anti)ferromagnetic phases [3,4,5].

Bismuth based compounds such as $BiFeO_3$, $BiMnO_3$, and $BiCrO_3$ with perovskite structure are promising multiferroics that are rather extensively studied [6-9]. $BiFeO_3$ for example, shows a ferroelectric ordering below 1123 K and a subsequent antiferromagnetic ordering below 643 K. However, because of spatially modulated spin structure, no macroscopic magnetization near room temperature has been observed [6]. A common byproduct in the non optimized solid state preparation of $BiFeO_3$ is the phase with chemical formula $Bi_xFe_yO_{1.5x+1.5y}$ particularly $Bi_2Fe_4O_9$ [10]. A unit cell of orthorhombic $Bi_2Fe_4O_9$ contains two formula units with two kinds of Fe atoms [11]. As shown in the inset of Figure 1a, there are four octahedral Fe ions on the sides of the cell and the remaining four tetrahedral Fe ions are in the interior. One can identify the super-exchange routes through which the tetrahedral Fe spins interact antiferromagnetically among themselves and with the octahedral Fe spins while there is a ferromagnetic coupling within a pair of octahedral spins (lying on the same side). These competing exchange interactions clearly generate spin frustration in this system which leads to a non collinear antiferromagnetic order in the *ab* plane [12]. Interestingly, $Bi_2Fe_4O_9$ has also been studied quite extensively over the past several decades for various functional applications such as a semiconductor gas sensor, and as a catalyst for ammonia oxidation [13]. In this letter, we report strong evidence for possible magneto-dielectric effects that extends close to room temperature in $Bi_2Fe_4O_9$.

Polycrystalline samples of $Bi_2Fe_4O_9$ were prepared by solid-state reaction at ambient pressure. High purity bismuth oxide $Bi_2O_3$ and iron oxide $Fe_2O_3$ were thoroughly mixed in stoichiometric proportion and were then compacted and calcinated at 850 °C for 12 hours in high purity alumina crucibles. The reacted product was ground, compacted, and reheated at 900 °C for 10 hours. Powder X- ray diffraction (XRD) data were collected using BRUKER D-8 advanced diffractometer (Cu $K_\alpha$ radiation, $\lambda$ = 1.5418 Å). Figure 1a shows the observed XRD pattern. All the major peaks of $Bi_2Fe_4O_9$ are identified. The sample exhibits nearly single phase orthorhombic structure with few impurity peaks corresponding to unreacted starting chemicals. The lattice parameters for $Bi_2Fe_4O_9$ are calculated to be a = 8.0627 Å, b = 8.5584 Å and c= 6.0086 Å. SEM and EDAX analysis were carried out using LEO 435 VP and Figure 1b shows SEM surface morphology of the sample. We find that the grains are well crystallized with size exceeding 6 μm in some cases. Some degree of porosity can be observed in the inset magnified image. The energy dispersive x-ray spectroscopic analysis at three different points in the sample approximately validated the 1: 2 ratio between Bi and Fe. This confirms the homogeneity and chemical composition of our samples.

Shown in Fig. 2 is the temperature dependence of magnetic susceptibility at an external field of 1 Tesla using a Quantum Design MPMS SQUID. The antiferromagnetic cusp at $T_N$ ~ 260 K is clearly observed. It is to be noted that room temperature resistivity of our sample is $2\times10^8$ ohm-cm which is relatively low. This is possibly caused by porosity as attested by the SEM data. The ferroelectric hysteresis was measured using a Radiant RT6000 test system. At room temperature no clear hysteretic behavior could be identified due to leakage current [14]. However, as the temperature was lowered to 250 K, the resistivity increased by two orders of magnitude and the observed P-E loop of the sample at 1 kHz is shown in Fig.3. We note that this loop is obtained at a temperature close to magnetic ordering. A differential scanning calorimeter scan (DSC) from 180 K to 470 K indicated a weak first order transition near the magnetic transition, as expected for a coupled magneto-electric material [15]. This is shown in the inset of Fig. 2. A sharp exothermic peak at 417 K was also observed but its origin could not be confirmed.

After establishing the antiferromagnetic and ferroelectric property, we turn to magneto-dielectric coupling of $Bi_2Fe_4O_9$. A standard procedure to confirm a coupling between the magnetic and electric order is an observation of inverse S-shaped anomaly in loss tangent across the magnetic transition as evidenced in $YMnO_3$ [16] and $ErMnO_3$ [17]. These anomalies in the loss factor and its reflection in dielectric constant can be altered by applying external perturbation such as magnetic field or stress. The temperature and magnetic field dependence of dielectric constant and loss ($\tan\delta$) at 123 Hz is shown in Figure 4. The dielectric constant ($\varepsilon$), plotted in the inset a, increases as temperature is increased and exhibits change of slope near $T_N$. In $\tan\delta$ this is seen as an appearance of broad shoulder. The first shoulder appears around 230 K and the second around 260 K. The shoulder appearing at 260 K corresponds to the AFM transition of $Bi_2Fe_4O_9$ and at 230 K it is related possibly to a *lock-in* transition at which modulation vector of AFM order of the $Fe^{3+}$ spins gets locked in a frustrated system [15]. We have also observed thermal hysteresis with increasing and decreasing temperatures below $T_N$. With the application of magnetic field, dielectric constant decreases and the loss factor increases and the inverse S-shape anomaly in $\tan\delta$ shifts to higher temperature. With 3 Tesla field, the peak shifts by 3 K. This provides clear evidence for strong coupling of electric polarization to the $Fe^{3+}$ magnetic moments.

While the comprehensive understanding of multiferrocity in orthorhombic $Bi_2Fe_4O_9$ would need a detailed study, it can be qualitatively understood in the following way. The induction of *improper* ferroelectricity in spin frustrated systems has attracted considerable attention in the recent past [18, 19]. The spin exchange interactions in $Bi_2Fe_4O_9$ are frustrated (see inset of figure 1a) and give rise to non collinear magnetization. Therefore the spins in the system can undergo a transition into an ordered antiferromagnetic state by releasing frustration through the lattice displacements. This provides a suitable mechanism for the magneto-dielectric effect in a frustrated spin system coupled with phonons. This picture is supported by the endothermic peak across the antiferromagnetic to paramagnetic transition as observed in DSC experiments. Furthermore, the neutron scattering experiments in $Bi_2Fe_4O_9$ clearly show a change in the lattice position parameters of the magnetic ions across the AF transition [12]. We note that the multiferrocity reported in the incommensurate spin frustrated system such as orthorhombic $TbMnO_3$ and $YMnO_3$ are in conformity with this picture [15, 20].

In conclusion, we have synthesized polycrystalline $Bi_2Fe_4O_9$ with sufficient purity and density that exhibits an antiferromagnetic ordering and ferroelectric hysteresis as revealed in the magnetic susceptibility and P-E loop respectively. We observe significant anomalies in dielectric constant and tan$\delta$ around the antiferromagnetic transition temperature $T_N \sim 260$ K that shifts to higher temperature with the application of magnetic field. This result unambiguously establishes the existence of coupling between electric and magnetic order in this industrial compound.

# Acknowledgement

We thank the Department of Science of Technology, Government of India, for the financial support under the FIST program to JNU, New Delhi and IIT, Kharagpur. AKS thanks CSIR, India for financial assistance.

**Figure Caption:**

Fig.1. (a) X-ray diffraction pattern of $Bi_2Fe_4O_9$ at room temperature. Plotted in y-axis is intensity in arbitrary units and all the major peaks are identified. The peaks marked as (*) belong to unreacted $Fe_2O_3$. The crystal structure is orthorhombic with space group *Pbam*. The inset shows $Fe^{3+}$ ion position in the ab plane of unit cell that contains two formula units. The symbol ● shows the two octahedral Fe atoms along c-axis which are ferromagnetically coupled and ○ shows the tetrahedral Fe atoms. The spin frustration is evident. The arrows mark the direction of spins. The thick, thin and dotted lines show strong, relatively weak and weakest AF coupling respectively. (b) SEM image of $Bi_2Fe_4O_9$. In the inset, a magnified version is shown. Some degree of porosity and voids can be seen.

Fig. 2. Temperature dependence of DC susceptibility ($\chi$) of $Bi_2Fe_4O_9$ is shown. The zero field cooled data is taken in the warming cycle with an external magnetic field of 1 T. The antiferromagnetic cusp at the Néel temperature is clearly observed. In the inset, normalized heat flow is plotted as a function of temperature. An endothermic peak reflecting a weak first order transition is identified around the magnetic transition.

Fig. 3. Polarization versus electric field loop of $Bi_2F_4O_9$ is plotted at 1 KHz frequency and T = 250 K.

Fig. 4. The loss factor tan$\delta$ is plotted as a function of temperature at 0T (■), 1T (○) and 3T (Δ) magnetic field. Inset a shows the corresponding variation in dielectric constant. Inset b shows the variation of tan$\delta$ at zero external field over a broad temperature range.

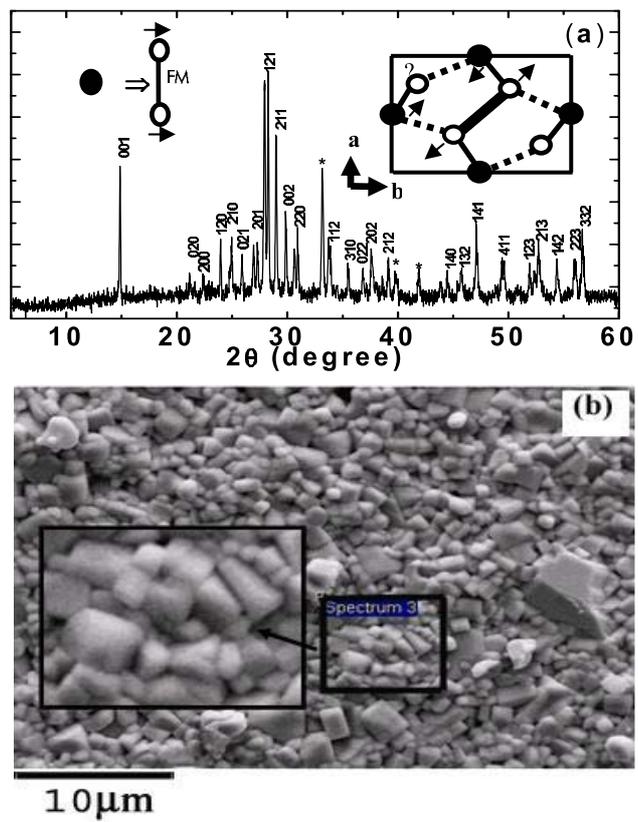

FIG.1.

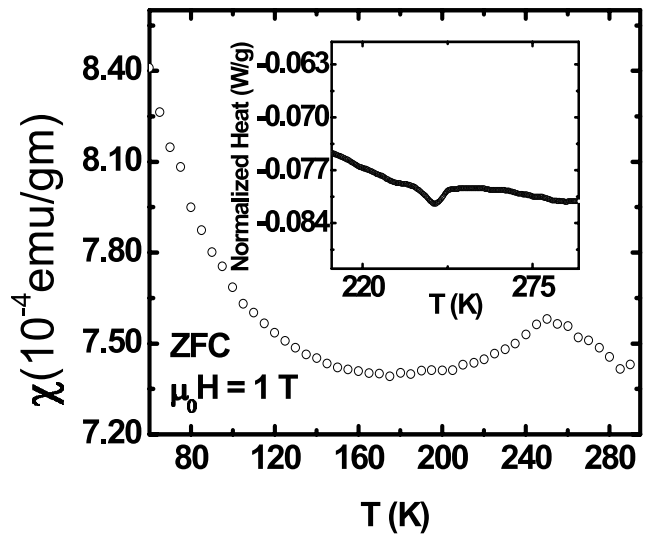

**FIG. 2.**

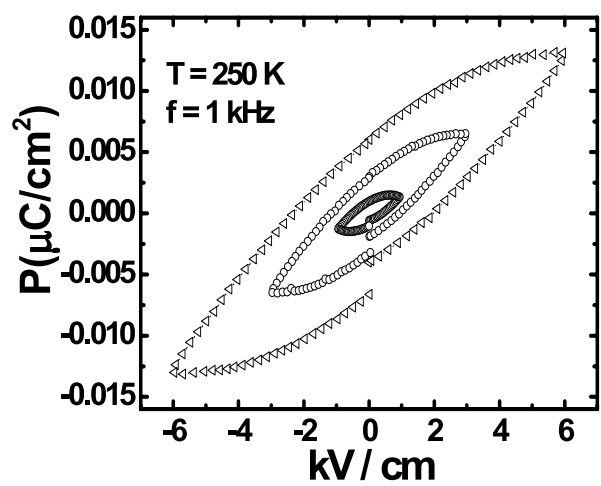

FIG. 3.

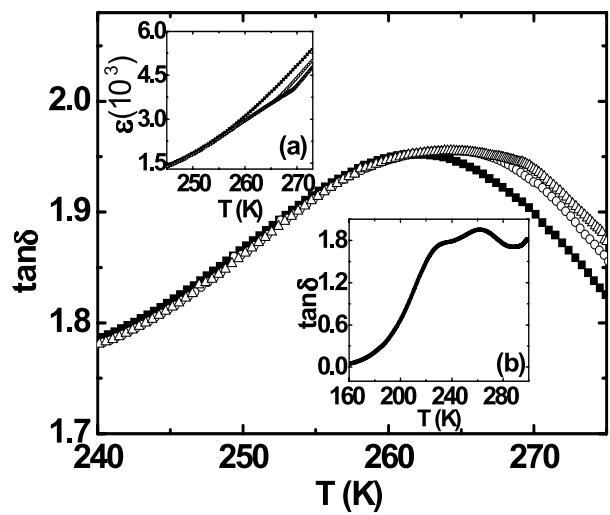

**FIG. 4.**